\documentclass[pra,twocolumn,showpacs]{revtex4}
\usepackage{graphicx}

\begin{document}

\title{Adiabatic state preparation in a cavity}
\author{J.~Larson}

\author{S.~Stenholm}
\affiliation{Physics Department\\
       Royal Institute of Technology (KTH)\\
       SCFAB,
       Roslagstullsbacken 21\\
       SE-10691 Stockholm,
       Sweden}



\begin{abstract}
The paper discusses the single-mode Jaynes-Cummings model
with time dependent parameters. Solvable models for two-level systems are
utilized to consider the changes in the photon distribution affected by the
passage of atoms through the cavity.  It is suggested that such systems may
be used as filters to modify the photon distribution. The effect can be
enhanced by repeatedly sending new atoms through the cavity. We show that
such filters can cut out either small or large photon numbers. It is also
shown that the method can be used to narrow down photon distributions and in
this way achieve highly non-classical sub-Poissonian states. Some
limitations and applications of the method are presented.
\end{abstract}

\maketitle

\section{Introduction}
\label{intro}

Cavity QED is based on the assumption that the mode selection allows us to
treat the radiation modes as a discrete set and consider states with only a
few modes excited. Thus the system is described by a finite number of
degrees of freedom; the rotating wave approximation (RWA) sets further
limits to the states which have to be included. In the case involving a
single mode only, we find the solvable Jaynes-Cummings model, which has
served as a prototype for many physical situations in Quantum Optics. 

 The Jaynes-Cummings model was introduced in 1963 \cite{jaynes-Cummings} and it has since played a central role in the development
of Quantum Optics; see the review \cite{shore-knight}. It has also served as
the theoretical foundation for analysing the results of experiments in
cavity QED \cite{walther}.

In all these applications, however, the model has been treated with constant
parameters. In our earlier work \cite{fredrik}, we discussed the model when
the parameters are allowed to change gradually during the interaction. We
now propose to continue our work on such time dependent phenomena in the
framework of the Jaynes-Cummings model.

In this paper, we assume that we can consider such a case where two-level
atoms consecutively interact with a single cavity mode sharp in frequency.
The various decay constants are taken to be zero. Thus all time scales
involved must be shorter than the damping times of the physical degrees of
freedom.

In Sec. \ref{jcmodel}, we introduce the basic Jaynes-Cummings model with time dependent
coefficients. We also discuss the possible solutions and the validity
conditions for the approach. In Sec. \ref{lz} we illustrate the filtering action
of atom dynamics by the simple results following from the Landau-Zener model.
The actual discussion of the photon state filtering is given in Sec. \ref{dk},
where we apply the solution of the more realistic Demkov-Kunike model. We
find that this may be utilised both to achieve sharpening of the photon
distribution and, if we prefer, as a low or high pass photon number
filter. With
numerical illustrations, we show that both narrowing of the state and
low pass
photon filtering may be achieved. In the former case, we show that highly
non-classical, sub-Poissonian photon distributions can be obtained. We also
present some approximate analytical expressions for the width of the ensuing
distributions. These are shown to provide god approximations to the
numerically obtained exact ones. Finally we conclude the paper in Sec. \ref{conc}
with a summary and some possible uses of these methods.

\section{Adiabatic Jaynes-Cummings model }\label{jcmodel}

In this paper, we consider the Jaynes-Cummings model with time dependent
parameters. There are two essential ones; within the RWA they are the cavity
detuning and the coupling constant. Both are to a large extent under
experimental control in modern experiments. Here we consider the situation
when these are made to change slowly during the course of the interaction
between the cavity and atoms. One assumption here is, of course, that the
timing of the atoms in the cavity can be controlled by the experiment. This
is non-trivial but not unrealistic. The change must be slow, because we want
the quantum state to follow the change adiabatically. This is
necessary in order to
preserve the identity of the single radiation mode involved; too fast a
change will mix in higher modes and destroy the simplicity of the system.
Thus we are working with an adiabatic extension of the ordinary cavity QED
system.

The Jaynes-Cummings Hamiltonian separates into two-component families of
coupled states, and we assume that this separation holds even when the
parameters are taken to vary slowly enough. Each family is characterised by
an initial photon number in the cavity, and the total solution emerges from
a sum over all initial photon states. The problem now becomes one of coupled
two-level systems with time dependent coefficients. Choosing these in
suitable ways, we can effect a variety of transformations on the initial
photon state. We suggest this as a tool to manipulate the photon
distribution and make a filter in the state space of the radiation. From
the literature, however, we know that two-level systems with time dependent
coefficients have played a central role in the understanding of quantum
phenomena. For a review of such solvable time dependent models see the review \cite{garraway}. Since a long time, many such models have been solved exactly and
in this paper we are illustrating the filtering property by an application
of these exact solutions.

The Hamiltonian we consider is the following one
\begin{equation}
\frac{\hat{H}}\hbar =\Omega \left( \hat{b}^{\dagger }\hat{b}+\frac 12\sigma
_3\right) +\frac{\Delta \omega }2\sigma _3+g\left( \hat{b}^{\dagger }\sigma
^{-}+\hat{b}\sigma ^{+}\right) ,  \label{a1}
\end{equation}
where the Pauli matrices are defined by
\begin{equation}
\begin{array}{lll}
\left[ \sigma _3,\sigma ^{\pm }\right]  & = & \pm 2\,\sigma ^{\pm } \\
&  &  \\
\sigma _{3}\mid \pm \rangle  & = & \pm \mid \pm \rangle .
\end{array}
\label{a2}
\end{equation}
The photon energy is given by $\hbar \Omega $ and the detuning is
defined as
\begin{equation}
\Delta \omega =\omega -\Omega ,  \label{a3}
\end{equation}
where $\hbar \omega $ is the energy separation in the two-level atom. The
first term in the Hamiltonian (\ref{a1}) is a constant of the motion, and we
can thus discuss the dynamics of the other terms neglecting this first one.
As usual, its effect can be added later. Here we assume that the parameters
are time dependent, $g(t)$ and $\Delta \omega (t)$. The latter is presumably
achieved by tuning the cavity frequency $\Omega (t)$.

We make an ansatz of the state in the form
\begin{equation}
\begin{array}{ccl}
|\Psi \rangle & = & c_0a_-(0)|0,-\rangle \\ \\ & & +\sum_{n=1}^\infty c_n\left[ a_{+}(n)\mid n-1,+\rangle
+a_{-}(n)\mid n,-\rangle \right] .  \label{a4}
\end{array}
\end{equation}
The initial states are assumed given by the coefficients $c_n$ and $a_{\pm
}^0(n).$ After the interaction period, these latter ones go over into $%
a_{\pm }^\infty (n).$

Without the first term of the Hamiltonian (\ref{a1}), we find for each family
the equations of motion
\begin{equation}
i\frac d{dt}\left[
\begin{array}{l}
a_{+}(n) \\
\\
a_{-}(n)
\end{array}
\right] =\left[
\begin{array}{lll}
\displaystyle{\frac{\Delta \omega }{2}} &  & g\sqrt{n} \\
&  &  \\
g\sqrt{n} &  & \displaystyle{-\frac{\Delta \omega }{2}}
\end{array}
\right] \left[
\begin{array}{l}
a_{+}(n) \\
\\
a_{-}(n)
\end{array}
\right] .  \label{a6}
\end{equation}
After a period of interaction, the solution of this problem can be written
in the form
\begin{equation}
\small
\left[
\begin{array}{l}
a_{+}^\infty (n) \\
\\
a_{-}^\infty (n)
\end{array}
\right] =\left[
\begin{array}{ccc}
\sqrt{w_n} &  & e^{-i\varphi _n}\sqrt{1-w_n} \\
&  &  \\
-e^{i\varphi _n}\sqrt{1-w_n} &  & \sqrt{w_n}
\end{array}
\right] \left[
\begin{array}{l}
a_{+}^0(n) \\
\\
a_{-}^0(n)
\end{array}
\right] .  \label{a7}
\end{equation}
If we now, after the interaction, perform a measurement to find the
two-level atom in one of the states $\mid \pm \rangle $, we find the cavity mode to
be in the corresponding state with the photon distribution
\begin{equation}
\begin{array}{lll}
P_n^{+} & = & \mid a_{+}^{\infty}(n+1)\mid ^2\mid c_{n+1}\mid ^2 \\
&  &  \\
P_n^{-} & = & \mid a_{-}^{\infty}(n)\mid ^2\mid c_n\mid ^2.
\end{array}
\label{a8}
\end{equation}
The asymmetry derives from our definition of the state (\ref{a4}), which
allows the components in Eq.(\ref{a7}) to be labelled with the same
$n.$ With the atom emerging in the upper state, we need to start with
one additional photon in the field. So the field distribution is modified by
$|a_{\pm}^{\infty}(n)|^2$, which we call filter functions.
Before the observation of the state of the two-level atom, the normalization
is given by
\begin{equation}\begin{array}{ccl}
1=\displaystyle{\sum_{n=0}^\infty} P_n & = & \displaystyle{\sum_{n=0}^\infty} \left[
  P_n^{+}+P_n^{-}\right] =a_{-}(0)\mid ^2\mid c_0\mid ^2\\ \\ & & +\displaystyle{\sum_{n=1}^\infty} \mid c_n\mid ^2\left[ \mid
a_{+}(n)\mid ^2+\mid a_{-}(n)\mid ^2\right].\end{array}  \label{a5}
\end{equation}
After the measurement of the two-level state, the probabilities (\ref{a8})
must be normalized to give
\begin{equation}
P^{\pm }=\sum_{n=0}^\infty P_n^{\pm }=1.  \label{a9}
\end{equation}

For general initial conditions, the solution (\ref{a7}) will depend on the
phase $\varphi _n$ which can be quite complicated. However, the treatment
simplifies if we restrict our attention to the two cases:
\begin{equation}
\begin{array}{lllll}
\mathrm{Case\,\, (a):} &  & |a_{-}^0|=1; &  &  a_{+}^0=0 \\
&  &  &  & \\
\mathrm{Case\,\, (b):} &  &  a_{-}^0=0, &  & |a_{+}^0|=1.
\end{array}
\label{a10}
\end{equation}
In these situations, no interference terms need to be considered. In
the following the initial values are taken to be as in the case (a). Since
the matrix in (\ref{a7}) is unitairy it follows that the filter functions remain normalized,
$|a_-^{\infty}(n)|^2+|a_+^{\infty}(n)|^2=1$. For the case (a),
$|a_-^{\infty}(0)|^2$ does not change and hence the normalization in
  (\ref{a5}) becomes $\sum_{n=0}^{\infty}|c_n|^2=1$.

 After the
observation of the two-level state, we have projected out a modified photon
state and applying a consecutive series of atoms with their initial states
according to (\ref{a10}), we can iterate the transformation (\ref{a8})
without having to consider the phase variable. This allows us to enhance the
filtering action exerted by a single atom affecting the cavity
mode. It is important to point out the fact that each time an upper
level atom is detected, and the photon distribution is modified by the
corresponding filter function, the photon numbers are shifted by one unit
as is seen in (\ref{a8}).

\section{The Landau-Zener model}\label{lz}

In the case solved by Landau and Zener, the parameters in Eq.(\ref{a6}) are
given by
\begin{equation}
\begin{array}{lll}
g= & g_0= & \mathrm{constant} \\
&  &  \\
\Delta \omega  & = & 2\lambda t.
\end{array}
\label{a11}
\end{equation}
The solution integrated over the time $(-\infty ,+\infty )$ is given by
\begin{equation}
\begin{array}{l}
w_n=\exp (-vn) \\
\\
v=\displaystyle{\frac{\pi g_0^2}{\lambda}} .
\end{array}
\label{a12}
\end{equation}
For the case (a) in (\ref{a10}) this gives
\begin{equation}
\begin{array}{l}
\mid a_{+}(n)\mid ^2=\left[ 1-\exp (-vn)\right]  \\
\\
\mid a_{-}(n)\mid ^2=\exp (-vn).
\end{array}
\label{a13}
\end{equation}
For the case (b) the roles of the two states are interchanged. Thus for
small values of the photon number, no transfer takes place, whereas for
large photon numbers, the atom approaches the adiabatic limit and the state
is transferred. The filter action is based on this property.

We can immediately see that the passage of a single two-level atom through
the cavity effects a filtering which eliminates the low photon number states
if the atom is found in its upper state and the high photon numbers if the
lower state is observed. The experimentalist has, of course, no control over
which is to be the case, but once the observation is done, the ensuing
cavity state is determined.

In order to observe the filtering action, we assume the cavity mode
initially to be in
a coherent state. Then we have
\begin{equation}
\mid c_n\mid ^2=\exp (-\bar{n})\frac{\bar{n}^n}{n!}.  \label{a14}
\end{equation}
From Eq.(\ref{a8}), we find that projecting on the lower state, we have the
normalized distribution
\begin{equation}
P_n^{-}=\exp (-\bar{n}e^{-v})\frac{\left( \bar{n}e^{-v}\right) ^n}{n!},
\label{a15}
\end{equation}
which is a Poisson distribution with the reduced average
\begin{equation}
\langle n\rangle_- =\exp (-v)\bar{n}\,.  \label{a16}
\end{equation}
Observing the other state, we find the distribution
\begin{equation}
P_{n-1}^{+}=c_N \left[ 1-\exp (-vn)\right] \exp (-\bar{n})\frac{\bar{n}^n%
}{n!},  \label{a17}
\end{equation}
which is the difference between the original photon distribution and the one
with the reduced average (\ref{a16}), $c_N$ is the normalization constant. This is clearly a method to filter out
the low photon numbers. The normalization coefficient can also easily be
determined from the relation
\begin{equation}
P^{+}=\sum_{n=0}^\infty P_n^{+}=c_N\displaystyle{\{}1-\exp \left[ \bar{n}(e^{-v}-1)\right]\displaystyle{\}} .
\label{a17a}
\end{equation}
After the detection of an upper level atom, the average photon number
is
\begin{equation}
\langle
n\rangle_+=\frac{1-\exp[\bar{n}(e^{-v}-1)-v]}{1-\exp[\bar{n}(e^{-v}-1)]}\bar{n}-1.
\end{equation}
If $\bar{n}\gg v$ the factor in front of $\bar{n}$ is approximately
unity and $\langle n\rangle\approx\bar{n}-1$, while if $v$ is large
the factor becomes greater that unity and we may have $\langle n\rangle>\bar{n}$.

If we repeatedly observe the lower state after a sequence of $m$ atoms
introduced in the lower level, we obtain the result
\begin{equation}
P_n^-(m)=\exp(-\bar{n}e^{-vm})\frac{(\bar{n}e^{-vm})^n}{n!},
  \label{a18}
\end{equation}
which will greatly enhance the filtering action. The corresponding
distribution for the upper level is
\begin{equation}
P_n^+(m)\propto\left[\prod_{\nu=1}^{m}|a_+^{\infty}(n+\nu)|^2\right]\exp(-\bar{n})\frac{\bar{n}^{n+m}}{(n+m)!}.
\end{equation} 
As each observation is totally random, the achievement of a
successful series of $m$ projections will become smaller and smaller.
However, the fact remains that after the observational sequence has been
recorded, the state of the cavity mode is known. This holds independently of
the actually observed sequence. If this contains alternating upper and lower
level observations, the appropriate transformation is to be taken from (\ref
{a13}).  As the subsequent
projections have to be normalized, the normalization will
contain the normalization coefficients of each preceding step in the
process. As long as each step involves a projection on to one of the atomic
states, no phase dependence needs to be taken into account.

In order to illustrate the filtering behaviour of the Landau-Zener model, we
show in figure 1, the filter function $\mid a_{-}^{\infty }(n)\mid ^{2m}$ for
the repeated number of atoms, $m=1,5$ and $25$ and $v=0.126$. For one atom we have the
simple result, which does not provide an efficient filtering, but with
increasing number of atoms, the high photon number parts become efficiently
suppressed, thus pushing the average to lower and lower values. This may, for example, be used
to cool an initially too hot thermal photon source. A
projection on the upper level efficiently selects the high photon
numbers instead. 
\begin{figure}[ht]
\centerline{\includegraphics[width=8cm]{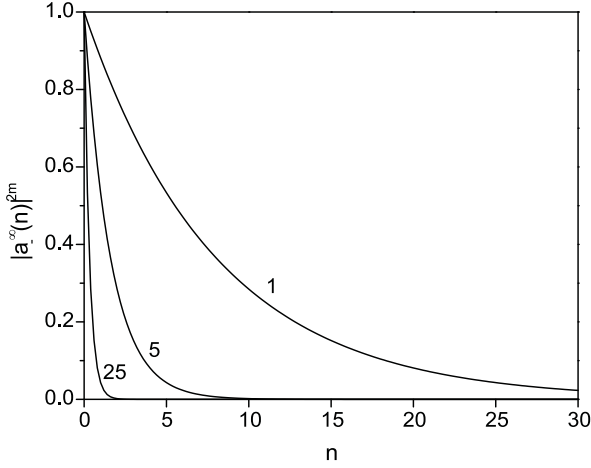}}
\caption {
This shows the filtering action achieved by the Landau-Zener
model, when the outgoing atom is projected on to its lower level. The curves
are presented for $v=\left( \frac{\pi g_0^2}\lambda \right) =0.126$. The three
curves give the result after the passage of one atom ($m=1)$, five atoms $%
(m=5$) and twenty-five atoms $(m=25)$, which is indicated in the figure. We can easily see the increased
efficiency of the filtering obtained.} 
\label{fig1}
\end{figure}
\section{The Demkov-Kunike model}\label{dk}

\subsection{Results of the model}

In order to simplify the discussion, we consider only the case (a) of Eq.(%
\ref{a10}). The parameters of the model  given in (\ref{a6}) are taken as
\begin{equation}
\begin{array}{lll}
\Delta \omega (t) & = & 2A_0\tanh \left( \frac tT\right)  \\
&  &  \\
g(t) & = & g_0\,\mathrm{sech}\left( \frac tT\right) .
\end{array}
\label{a19}
\end{equation}
This is a three parameter model, which integrated over the time $(-\infty
,+\infty )$ gives the solution
\begin{equation}
\begin{array}{lll}
\mid a_{-}^\infty (n)\mid ^2 & = & \displaystyle{\frac{\cos ^2\left( \pi T\sqrt{\left(
g_0^2n-A_0^2\right) }\right) }{\cosh ^2(\pi TA_0)}} \\
&  &  \\
\mid a_{+}^\infty (n)\mid ^2 & = & 1-\mid a_{-}^\infty (n)\mid ^2.
\end{array}
\label{a20}
\end{equation}
For photon numbers such that $g_0^2n<A_0^2,$ we have
\begin{equation}
\mid a_{-}^\infty(n) \mid ^2=\frac{\cosh ^2\left( \pi T\sqrt{\left(
A_0^2-g_0^2n\right) }\right) }{\cosh ^2(\pi
TA_0)}\,{\rightarrow}\, 1,  \label{a21}
\end{equation}
as $n\rightarrow 0$. This must hold for all models as long as
we consider the case (a) of Eq. (\ref{a10}). The filter function provided by this model, thus starts from unity, and
within a range of $(A_0/g_0)^2$ of photon numbers goes to an oscillating
function with the amplitude $\mathrm{sech}^2(\pi TA_0)$. For small coupling
constants $g_0$ , this is thus essentially a low photon number filter like
the Landau-Zener case.

For a strong coupling case, $g_0>A_0$, the function is essentially an
oscillating one, with the first maximum of $|a_-^{\infty}(n)|^2$ at
\begin{equation}
n_M=\frac{\left( 1+T^2A_0^2\right) }{T^2g_0^2}\sim \frac 1{T^2g_0^2}.
\label{a22}
\end{equation}
Even if the distance to the surrounding zeros is of the same order of
magnitude, the trigonometric dependence on $n$ provides a rather sharp
cut-off around the maximum. Application of the transformation (\ref{a21}) to
a coherent state with most probable photon number around $n_M$ will result
in a considerable narrowing of the photon distribution. It is advantageous
to choose the parameter $A_0$ as small as possible, because then the
oscillation amplitude $\mathrm{sech}^2(\pi TA_0)$ is maximized. The same happens
when $T$ is chosen short, however, then the oscillating period becomes large. We thus want to work close to the non-adiabatic
limit, of the two-level model, in this case.

In the non-adiabatic limit, the projection on the lower state of the atom
gives the filtering
\begin{equation}\begin{array}{ccl}
\mid a_{-}^\infty(n) \mid ^2 & = & \cos ^2\left( \pi T\sqrt{\left(
g_0^2n-A_0^2\right) }\right)\mathrm{sech}^2(\pi TA_0) \\ \\ & \approx & \cos
^2\left( \pi T\sqrt{g_0^2n}\right) ,  \label{a23}
\end{array}
\end{equation}
which gives a simple filtering around the maxima
\begin{equation}
n_M=\frac{ k^2}{T^2g_0^2}.\,\,\,\,\mathrm{ (k=0,1,2...)}
\label{a24}
\end{equation}
The zeros of the distribution are given by the expression
\begin{equation}
n_0=\frac{\left(k+\frac{1}{2}\right)^2}{T^2g_0^2}  \label{a24a}
\end{equation}
The width of the distribution is hence approximated by
\begin{equation}
\Delta n\sim n_0-n_M\sim \frac k{T^2g_0^2}.  \label{a24b}
\end{equation}
For a Poisson distribution, $\Delta n\sim \sqrt{\bar{n}}.$ Thus to set the maximum $n_M$
to a large value  requires a large $k$. The width (\ref{a24b}) then scales
as the square root of $\bar{n}$ with the implication that it matters little,
which peak is chosen for the filtering. The factor $Tg_0$ can well be chosen
close to unity in this parameter range.  One may think that for optimal
filtering action, the first maximum should be chosen.  However, this does
not exclude large values of $n_M.$ By a suitable choice of $T,$ we can make
both $A_0T$ and $Tg_0$ small even if $g_0>A_0$. Thus the three parameters
may be utilised to achieve desired filtering operation.

By sending atoms repeatedly through the cavity, in the case of a sequence of
$m$ recorded lower level observations, we obtain the filtering
\begin{equation}
P_{n}^{-}(m)=\cos^{2m}\left( \pi T\sqrt{g_0^2n}\right) \mid c_n\mid ^2.
\label{a25}
\end{equation}
The powers of the trigonometric function becomes rapidly sharper with
increasing $m.$ Thus the filtering is fast and efficient. As the repeated
application of the filtering process sharpens up the photon distribution, we
expect the states to rapidly become non-classical. In order to explore this,
we have investigated the Q-parameter, see \cite{Mandl},
\begin{equation}
Q=\frac{\langle n^2\rangle -\langle n\rangle ^2-\langle n\rangle }{\langle
n\rangle },  \label{a26}
\end{equation}
as a function of the coupling strength. For a range of initial coherent
states, we find indeed negative values; see the next section.

In the adiabatic limit, $T$ growing large (i.e. $A_0T$ being large), the oscillational amplitude
becomes small, and we expect the filtering process to become less efficient
\begin{equation}
\mathrm{sech}^2(\pi TA_0)\sim 4\exp (-2\pi TA_0)\ll 1.  \label{a27}
\end{equation}
Then $\mid a_{-}^\infty(n) \mid ^2$ from (\ref{a20}) decreases. In this case, we
may also choose the weak coupling situation $g_0<A_0$ and write
\begin{equation}
\begin{array}{lll}
P_n^{-}(m) & = & \left( \displaystyle{\frac{\cosh ^2\left( \pi T\sqrt{\left(
A_0^2-g_0^2n\right) }\right) }{\cosh ^2(\pi A_0T) }}\right) ^m\mid c_n\mid ^2 \\
&  &  \\
& \approx  & \left\{ \exp \left[ 2\pi T\left( \sqrt{\left(
A_0^2-g_0^2n\right) }-A_0\right) \right] \right\} ^m\mid c_n\mid ^2 \\
&  &  \\
& \approx  & \exp \left( -\pi \frac{g_0^2Tm}{A_0}n\right) \mid c_n\mid ^2.
\end{array}
\label{a28}
\end{equation}
This form simplifies some of the arguments below.

\subsection{Numerical results}\label{num} 
In this section we illustrate our argument by some numerical
examples. As all relevant parameters has the dimension of time or
frequency, fixing one time scale allows us to use dimensionless units
in the calculations.
\begin{figure}[ht]
\centerline{\includegraphics[width=8cm]{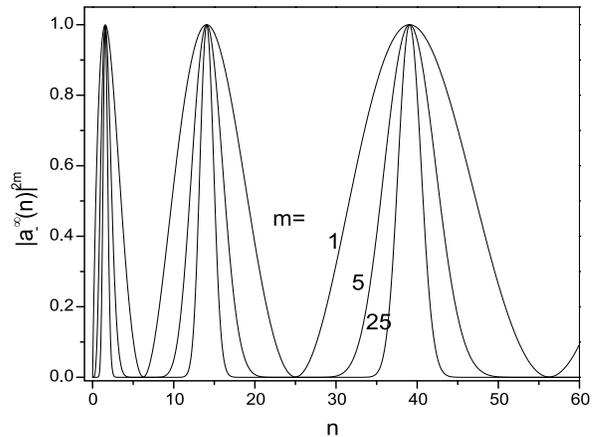}}
\caption[figure2] {\label{fig2}
This figure shows the  filter functions for
the Demkov-Kunike model. In our dimensionless units we have the parameters, $A_0=T=0.1$ and $g_0=4$
. Here we also look at one, five and twenty-five atoms passing the cavity ($%
m=1,5,25).$ In (a) the lower level filter function
$|a_-^{\infty}(n)|^{2m}$ is plotted. The increase of atom number rapidly narrows the distribution,
and we can also see how the width scales approximately as the position of
the peak of the function. This allows us to use the filter on a coherent
state distribution situated at any peak of our choice. The filter
function $\prod_{\nu=1}^{m}|a_+^{\infty}(n+\nu)|^2$ is shown in
(b). The maxima in (b) are shifted to the left by approximately $m/2$ units. The
numbers in the figures are again indicating how many atoms $m$ have been detected.  
}

\end{figure}

As we have seen, if we repeatedly project on the lower or the upper atomic
level, the new photon distribution will be modified by
$|a_+^{\infty}(n)|^{2m}$ or $\prod_{\nu=1}^m|a_+^{\infty}(n+\nu)|^2$
respectively. In the latter case the initial photon distribution will
also be shifted by $m$ units. In figure 2 (a) and (b) we plot the two
functions above for $m=1,5$ and $25$. The maxima in (b) are shifted to the
left by $m/2$ units. For small $A_0T$ the function $|a_-^{\infty}(n)|^{2m}$ will be approximately
zero everywhere except where $n\approx n_M$ and $n_M$ is close to the maxima
(\ref{a24}). If the initial photon distribution, given
by $|c_n|^2$, has a maximum near one
$n_M$ and a width smaller than the period of the filter function, then
we expect a sharp final distribution if a series of $m$ lower level
atoms is detected. We know that the
$Q$-parameter defined in (\ref{a26}) is positive for a
super-Poissonian and negative for sub-Possonian state, which signals
non-classical light. A plot of $Q$
as a function of $g_0$ is shown in figure 3 for an initial coherent
 state with $\bar{n}=100$, the number of atoms $m=25$ and
$T=A_0=0.1$. We see that the field distribution $P_n^-$ is
sub-Poissonian for a large range of $g_0$:s. Note that one could use
the $Q$-parameter to find the optimal parameters $T$ and $g_0$ in
order to obtain the most non-classical light.
\begin{figure}[ht]
\centerline{\includegraphics[width=8cm]{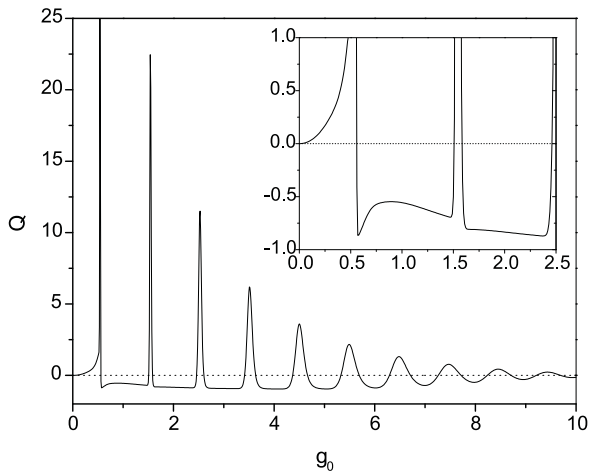}}
\caption {This shows the variation of the Q-parameter with coupling
strength. We must have $Q\geq -1,$ but all negative values indicate
non-classical photon statistics. The
intial state assumed here has average photon number  $\overline{n}=100;$ the
interaction is defined by $T=A_0=0.1.$ In this figure, we consider the case
with $m=25$ atoms having successfully passed the cavity and been found
in their lower level. The insert zooms
in on the first non-classical regions, and shows the degree of
non-classicality which is achieved. The limit for classical behaviour, $Q=0$
, is indicated by the dotted line.}
\label{fig3}
\end{figure}
As noted in the previous section, since the width of the maxima of the filter
function goes as $\sqrt{n_M}$, just like the width of a coherent state,
the ratio between them is independent of $n$. So if we know
$g_0$, $T$ and $A_0$, the filtering effect is chiefly independent of which
maximum $n_M$ the initial coherent state is centred at. The distribution
$P_n^-$ for an initially coherent state, centred around the third
maximum $\bar{n}=25$ of the filter function, is shown in figure 4 with $T=A_0=0.1$,
$g_0=4$ and the number of atoms $m=1,5$ and $25$.
\begin{figure}[ht]
\centerline{\includegraphics[width=8cm]{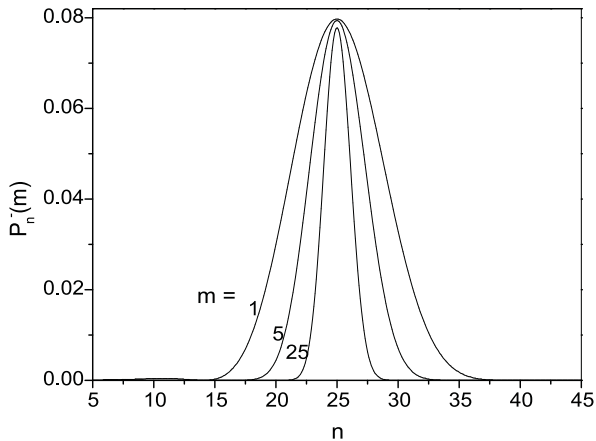}}
\caption {
This figure shows the narrowing of the phton distribution actually
achieved with one, five or twenty-five atoms ($m=1,5,25)$ acting on a
coherent state centered on $\overline{n}=25$, which in this case is the
second maximum; c.f. figure 2 (a). The dimensionless parameters are $T=A_0=0.1$ and
$g_0=4.$ 
}
\label{fig4}
\end{figure}

In order to have a low photon number filter it was found to be advantageous to use a large
adiabaticity parameter $\Lambda=A_0T$ and a small coupling constant
$g_0<A_0$. A large adiabaticity parameter means that the oscillating
part of the filter function is small and the weak coupling implies
that the hyperbolic behaviour in the filter function extends over a
long distance. In figure 5 the filter
function $|a_-^{\infty}(n)|^2$ is plotted for the parameters
$g_0=0.2$, $T=0.9$ and $A_0=0.6$. Except for the oscillations, this
should be compared with the case $m=1$ in figure 1. 
\begin{figure}[ht]
\centerline{\includegraphics[width=8cm]{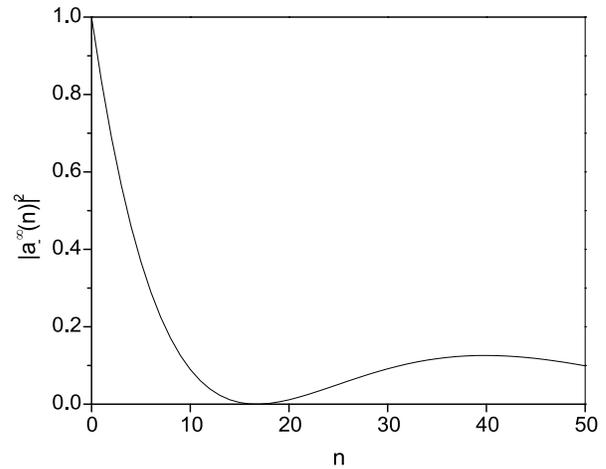}}
\caption {
This is the filtering function after the passage of one atom in
the limit when the Demkov-Kunike model acts as a low photon number filter.
Here $T=0.9,$ $A_0=0.6$ and $g_0=0.2$. Except for the oscillations, this is
similar to the Landau-Zener result, figure 1 ($m=1$).
} 
\label{fig5}
\end{figure}

\subsection{Analytical approximations}\label{anapp}
We have considered two limits of the two-level model: The
non-adiabatic one which could give a sharpened photon distribution and the
adiabatic one which is used to filter out the high or the low photon number states.
In this section we derive analytic expressions for the widths of the
sharpened distribution and the low photon number filter.

The filter function is given by
Eq. (\ref{a23}) in the non-adiabatic limit, and its maximum by the
result (\ref{a24}). After the passage of
$m$ atoms, all of them detected in their lower level, we consider the photon distribution. The FWHM,
$\Delta n_a$, of the filtering function  is determined from
\begin{equation}\label{eq:fwhma1}
\begin{array}{c}
\cos^{2m}\left(\pi
Tg_0\sqrt{n_M\pm\Delta n_a/2}\right)=\,\,\,\,\,\,\,\,\,  \\ \\ \,\,\,\,\,\,\,\,\cos^{2m}\left(n_M\pm\arccos\left(0.5^{(1/2m)}\right)\right),
\end{array}
\end{equation}
where the signs $\pm$ derive from the fact that the filtering function is
not symmetric around $n_M$ due to the $\sqrt{n}$ - behavior. Equation
(\ref{eq:fwhma1}) determines two different $\Delta n_a$ and to get the
appropiate value we take their average
\begin{equation}
\Delta n_a=\frac{4x_m\sqrt{n_M}}{\pi Tg_0},
\end{equation}
where $x_m=\arccos\left(\sqrt[2m]{0.5}\right)$ which starts at $\pi/4$ and falls to
zero for increasing $m$. For an initial Poission
distribution, we have for the half-width to a good approximation
$\Delta n_p=\sqrt{8n_M\ln(2)}$, and the width of the final
distribution $P_n^-(m)$ is approximately
\begin{equation}\label{FWHM}
\Delta n^2=\frac{\Delta n_a^2\Delta n_p^2}{\Delta n_a^2+\Delta
  n_p^2}=\frac{16x_m^2n_M\ln(2)}{\pi^2T^2g_0^2\ln(2)+2x_m^2}.
\end{equation}
For a narrow $\Delta n_a$ we have $\Delta n\approx\Delta
n_a\sim x_m$. A plot of $\Delta n$  for $g_0=4.5$,
$A_0=T=0.1$ and $\bar{n}=19.8$ as a function of $m$ is shown in figure 6 together with
exact numerical results for the FWHM. 
\begin{figure}[ht]
\centerline{\includegraphics[width=8cm]{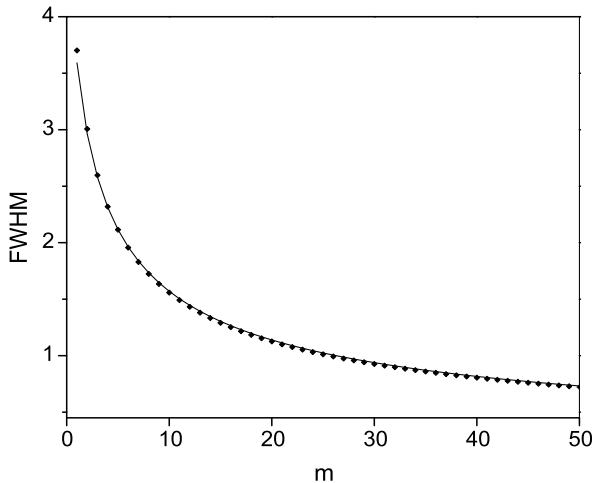}}
\caption {
This figure compares the approximate expression (\ref{FWHM}) (solid line) for
the width of the final photon distribution with the exact one obtained
numerically (diamonds). The dimensionless parameters are here $T=A_0=0.1$ and $g_0=4.5$.
The plot shows the influence of increasing atom number $m$ on an initially
coherent state with $\overline{n}=19.8.$
}
\label{fig6}
\end{figure}

Now we turn to the adiabatic limit and the low photon number filter. With the right
parameters it is possible to achieve such filtering by repeatedly projecting
on the lower atomic level. The width, $\Delta n$ at $e^{-1}$, of
this low photon number filter is derived from 
\begin{equation}\label{eq:exactwidth}
\cosh^{2m}\left(\pi T\sqrt{A_0^2-g_0^2\Delta n}\right)\mathrm{sech}^{2m}(\pi
  TA_0)=e^{-1}
\end{equation}
which determines $\Delta n$. The result is complicated and not very
informative. However
if we instead use the fact that we are in the adiabatic regime and take
the last expression in Eq. (\ref{a28}) as the filter function, we obtain the
width
\begin{equation}\label{eq:appwidth}
\Delta n=\frac{A_0}{\pi g_0^2Tm}.
\end{equation}
The width clearly decreases as $m$ becomes large, so that the high
photon numbers are effectively filtered out.
In Fig. 7 the width (\ref{eq:appwidth}) is plotted together
with the exact width obtained from Eq. (\ref{eq:exactwidth}) as a function of
$m$ for the parameters $A_0=2$ and $T=g_0=1$. Again we have a very
good agreement. 
\begin{figure}[ht]
\centerline{\includegraphics[width=8cm]{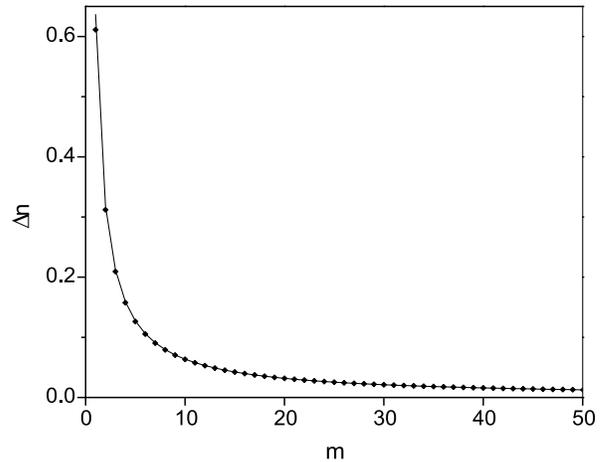}}
\caption {This figure compares the approximate expression (\ref{eq:appwidth}) (solid) for the width
of the final photon distribution with the exact one obtained from (\ref{eq:exactwidth})
(diamonds).
Here we are in the limit where the interaction acts as a low photon
filter and we see the influence
of increasing atom number $m$.
The dimensionless parameters are $A_0=2$ and $T=g_0=1$. As we have a low photon
filter, this may be interpreted as the range over which the filter acts.
Alternatively we may use it as a high photon one with the corresponding
interpretation.
}
\label{fig7}
\end{figure}

\section{Conclusion}\label{conc}

In this paper, we consider the situation where individual atoms are made to
traverse a cavity at the will of the experimenter. During the passage of the
atom through the cavity, the parameters of the model are made to vary in
order to achieve a desired effect. This variation must be slow enough to
preserve the energy levels of the interacting systems intact. For slow atoms
this should present no problem, but the time intervals involved are limited
to those allowed by the various relaxation processes which are omitted in
our treatment. When actual experimental parameters are introduced, the
influence of relaxation processes can easily be included by performing
numerical integration of the time evolution. Here we want to present the
main ideas, and such detailed investigations have been excluded.

In order to illustrate our ideas, we choose to utilise the existence of
analytically solvable models for the interactions. Choosing the simple
Landau-Zener model as a prototype, we then investigate in detail the
behaviour for the related but much more realistic Demkov-Kunike model. They
both allow filtering action for either low or high photon numbers. If low
photon numbers are favoured, the process can be used to cool the photon
distribution, but favouring high photon numbers we can sharpen the
distribution or heat it up. For sharpening it, the Demkov-Kunike model
provides a much more efficient tool, which we investigate in some detail. In
particular, we show that strongly sub-Poissonian distributions can be
achieved.

There are many treatments of time-dependent two-level systems in the
literature. Many cases have been solved analytically and the standard
method is to reduce the problem to a differential equation with a
known solution \cite{shore}. Such models have been utilized widely
to model time-dependent atomic processes \cite{nikitin}. Any one of
these could be used in our present scheme to analyze the result of the
atom-cavity interaction. However, as we are here interested in the
modification of the photon distribution, these earlier works are only
of interest so far as they provide novel filter functions for the
quantum field. Consequently we have here choosen to investigate two of
the best known models only; they provide enough insight into the
possibilties and limitations of our suggested approach. More intricate
technical details of the actual solution are of little interest in the
present context.

In order to obtain the most dramatic effects, we have to let several atoms
pass through the cavity. Their state is determined experimentally after the
interaction period, and the corresponding modification of the photon
distribution becomes known. This occurs for either of the two possible
outcomes of the experiment. However, if we want to achieve a pre-assigned
series of observations, this can only be found with a certain probability.
Several attempts may be necessary to observe the desired sequence. However,
once it has been found, the cavity state is also known, and it may be used
in any subsequent experiment within the period allowed by the cavity
relaxation rates.

Our suggested approach is, of course, part of the broad range of works
called ''state preparation'' and ''state engineering''. Much work has been
done in this area, and we cannot possibly relate our effort to all of these.
For an overview consult e.g. the special issue \cite{stateprep} on Quantum State
Preparation and Measurement. Here we only want to point out, that the
modification of the state is determined by the dynamic interaction between
the atoms and the cavity mode. The repeated observation leading to
projection onto the new cavity state does not, in itself, shape the
distribution, it only picks one out of two possibilities. 

There are many cavity QED schemes where the field in the cavity is
modified by transit of atoms through the cavity, see
e.g. \cite{haroche} and \cite{walther2} and refereces therein. Methods
for prepering various states of the cavity field have been proposed, e.g.
Fock states \cite{walther3}, Schr\"odinger cat states
\cite{haroche} and EPR states between two cavity modes \cite{haroche2}. In 
these works, the time dependent feature is the transit of the atom
through the cavity, and a suitable measurement at the end of the
interaction probes the result achieved. The idea is, of course,
closely related to our approach, but we have added the novel feature
that the parameters of the coupled atom-cavity system are deliberately
changed so that we achieve a desired effect. Thus the modification of
the quantum state is effected by the control imposed by the
experimenter and not only by the interaction followed by a detection procedeure.

We have utilized solvable models to illustrate out basic ideas. These make
it possible to assert the influence of the various parameters in a
straightforward way. In an actual experimental situation, we may, of course,
perform the corresponding dynamical calculations on a computer, which allows
one to introduce various effects related to experimental conditions and
non-ideal observational methods. Such considerations are, however, best
performed in connection with realistic laboratory situations.


\newpage

\begin{thebibliography}{12}

\bibitem{jaynes-Cummings} E. T. Jaynes and F. W. Cummings, Proc. IEEE 
\textbf{51}, 89 (1963).

\bibitem{shore-knight} B. W. Shore and P. L. Knight J. Mod. Opt. \textbf{40}, 1195 (1993).

\bibitem{walther} G. Raithel, Ch. Wagner, H. Walther, L. M. Narducci and M.
O. Scully, 1994,  \textit{Cavity Quantum Electrodynamics, Advances in Atomic,
molecular and Optical Physics, Supplement 2, }, editor P. R. Berman
(Academic Press), pp 57-122. 

\bibitem{fredrik} F. Mattinson, M. Kira and S. Stenholm, J. Mod. Opt. 
\textbf{48}, 889 (2001).

\bibitem{garraway} B. M. Garraway and K.-A. Suominen, Rep. Progr. Phys. 
\textbf{58}, 365 (1995).

\bibitem{Mandl} L. Mandel and E. Wolf, 1995, \textit{Optical Coherence
    and Quantum Optics}, (Cambridge University Press), p 627.

\bibitem{shore} B. W. Shore, \textit{The Theory of Coherent Atomic
    Excitation}, (John Wiley \& sons, New York 1990), sections 5.5 and 5.6.

\bibitem{nikitin} E. E. Nikitin and S. Ya. Umenskii, \textit{Theory of
    Slow Atomic Collisions}, (Springer Verlag, Heidelberg
    1999), sections 7-9.

\bibitem{stateprep} W. P. Schleich and M. G. Raymer, editors,
  J. Mod. Opt. \textbf{44}, 2021 (1997). 

\bibitem{haroche} J. M. Raimond and S. Haroche, Atoms and
    Cavities: The birth of a Schr\"odinger cat of the radiation field,
    in \textit{International Trends in Optics and Photonics},
    ed. T. Asakura (Springer Verlag, Heidelberg 1999), pp 40-53.

\bibitem{walther2} S. Brattke, B. T. H. Varcoe and H. Walther,
  Phys. Rev. Lett. \textbf{86}, 3534 (2001).

\bibitem{walther3} M. Franca Santos, E. Solano and R. L. de Matos
  Filho, Phys. Rev. Lett. \textbf{87}, 093601 (2001); S. Brattke,
  B. T. H. Varcoe and S. Walther, Opt. Express \textbf{8}, 131 (2001);
  P. Bertet, S. Osnaghi, P. Milman, A. Auffeves, P. Maioli, M. Brune,
  J. M. Raimond and S. Haroche, Phys. Rev. Lett. \textbf{88}, 143601
  (2002).

\bibitem{haroche2} A. Rauschenbeutel, P. Bertet, S. Osnaghi,
  G. Nogues, M. Brune, J. M. Raimond and S. Haroche, Phys. Rev. A
  \textbf{64} 050301 (2001).

\end{thebibliography}
\end{document}